\documentclass[letterpaper,twocolumn,10pt]{article}

\usepackage{usenix-2020-09}
\usepackage{tikz}
\usepackage{amsmath}
\usepackage{subcaption}
\usepackage{balance}
\usepackage{float}
\usepackage{booktabs}
\usepackage{listings}
\usepackage{xcolor}
\usepackage[backend=biber,style=numeric]{biblatex}
\begin{document}

\pagestyle{empty}

\title{\Large \bf Exploring and Exploiting the Resource Isolation Attack Surface of \\
WebAssembly Containers}

\author{
{\rm Zhaofeng Yu$^{1}$},\,
{\rm Dongyang Zhan$^{1}$\thanks{\,Corresponding author: Dongyang Zhan (Email: zhandy@hit.edu.cn)}},\,
{\rm Lin Ye$^{1}$},\,
{\rm Haining Yu$^{1}$},\,
{\rm Hongli Zhang$^{1}$},\,
{\rm Zhihong Tian$^{2}$}\\
$^1$Harbin Institute of Technology,\quad  $^2$Guangzhou University\\
{\tt yuzhaofeng@stu.hit.edu.cn} \\
{\tt \{zhandy, hityelin, yuhaining, zhanghongli\}@hit.edu.cn} \\
{\tt tianzhihong@gzhu.edu.cn}
}

\maketitle

\begin{abstract}

Recently, the WebAssembly (or Wasm) technology has been rapidly evolving, with many runtimes actively under development, providing cross-platform secure sandboxes for Wasm modules to run as portable containers. Compared with Docker, which isolates applications at the operating system level, Wasm runtimes provide more security mechanisms, such as linear memory, type checking, and protected call stacks. Although Wasm is designed with security in mind and considered to be a more secure container runtime, various security challenges have arisen, and researchers have focused on the security of Wasm runtimes, such as discovering vulnerabilities or proposing new security mechanisms to achieve robust isolation. However, we have observed that the resource isolation is not well protected by the current Wasm runtimes, and attackers can exhaust the host's resources to interfere with the execution of other container instances by exploiting the WASI/WASIX interfaces. And the attack surface has not been well explored and measured. In this paper, we explore the resource isolation attack surface of Wasm runtimes systematically by proposing several static Wasm runtime analysis approaches. Based on the analysis results, we propose several exploitation strategies to break the resource isolation of Wasm runtimes. The experimental results show that malicious Wasm instances can not only consume large amounts of system resources on their own but also introduce high workloads into other components of the underlying operating system, leading to a substantial performance degradation of the whole system. In addition, the mitigation approaches have also been discussed.
\end{abstract}

\section{Introduction}

Virtualization technologies are continuously evolving, progressing from traditional virtual machines to Linux containers, and now to WebAssembly (Wasm)\cite{webassembly}. These emerging technologies offer enhanced usability, such as faster, more flexible deployment capabilities and more efficient resource utilization. However, they introduce new security challenges. For example, Docker containers, by sharing the Linux kernel, enable more rapid and flexible deployment compared to traditional virtual machines but also expand the attack surface. This shared kernel architecture makes it possible for attackers to compromise the host operating system, thereby affecting both the host machine and other container instances\cite{ghavamnia2020confine,xing2023hybrid,zhan2022shrinking}. As the next generation of virtualization technology, Wasm offers a platform-independent solution that allows applications to be deployed more flexibly across servers, cloud environments, or even within browsers, while achieving near-native execution speed\cite{bosamiya2022provably,johnson2023wave,menetrey2023comprehensive,haas2017bringing}. However, Wasm also introduces its own set of vulnerabilities, creating new attack surfaces that could potentially threaten the host environment.

Wasm is a portable binary instruction format that is independent of language, hardware, and platform, and it can serve as a compilation target for other programming languages. Developers can write applications in various languages and then compile them into portable Wasm modules, which can be executed as Wasm instances via Wasm runtimes. Initially, Wasm was designed for deploying web applications within browsers. In recent years, it has been adopted in a broader range of domains, such as the IoT\cite{ribeiro2024wasmico,kuribayashi2023dynamic,pham2023webassembly} and cloud-edge computing\cite{sekigawa2023web,lim2023advanced,kakati2023webassembly}, where its versatility and efficiency have been widely recognized.

The key to Wasm's cross-platform capability lies in a standardized set of system interfaces. As an instruction format, Wasm itself does not provide direct methods to access operating system functionalities. To enable general applications to function properly within a Wasm runtime (e.g., accessing the file system), a standardized set of WebAssembly System Interfaces (WASI)\cite{WASI} is defined. Through calls to the WASI interfaces implemented by the Wasm runtime\cite{Wasmtime,Wasmer,WasmEdge}, the code within Wasm modules can access part of operating system features. Currently, there is also another set of interfaces called WASIX\cite{WASIX}. WASIX is an extension of WASI, incorporating additional APIs to offer richer functionality, allowing it to support more complex applications.

As the intermediate layer between the Wasm instances and the underlying operating system, the WASI/WASIX interfaces not only offer access to system resources but also restrict how Wasm instances interact with the operating system. Wasm runtimes typically embed strict validation logic when implementing these interfaces, enforcing isolation between Wasm instances and the underlying system. For example, when a Wasm instance attempts to access the file system, the Wasm runtime will check the target file path and restrict the module to accessing only files within specified directories. This approach ensures that Wasm instances operate within a controlled and secure environment, mitigating potential security risks while maintaining the flexibility and portability that Wasm offers.

However, there are some security problems in the Wasm runtimes that prevent them from fully isolating Wasm instances. Based on our observations, several of the most widely used Wasm runtimes impose insufficient resource isolation. Through various experiments, we find that Wasm instances can exploit these interfaces to cause significant damage on the underlying operating system or other Wasm instances. For example, a Wasm instance can exhaust system resources through exploiting file-related interfaces. These findings indicate that WASI/WASIX interfaces are not stringent enough to fully isolate Wasm instances. In recent years, many studies have emerged focusing on the Wasm security, but most of them focus on either the vulnerabilities and attacks within Wasm\cite{romano2021empirical,lehmann2020everything,lehmann2020everything,romano2022wobfuscator,rokicki2022port,oz2023rob,katzman2023gates,puddu2024lack} or the design of more secure mechanisms to improve its security\cite{narayan2021swivel,bosamiya2022provably,johnson2023wave,lei2023put,yavarzadeh2023half,menetrey2023comprehensive}. However, the attack surface of Wasm runtimes has not been systematically measured, and how attackers can perform resource-related attacks through the attack surface has not been studied.

Currently, resource availability issues in virtualization environments have received widespread attention. Prior works \cite{sharma2016containers,gao2019houdini,yang2021demons,xiao2023attacks} have investigated such issues in container and virtual machine environments, showing that attackers can bypass isolation boundaries to exhaust host resources and launch denial-of-service attacks. In this paper, we focus on resource availability in the context of Wasm runtimes. Specifically, we explore the resource isolation attack surface of Wasm runtimes and analyze how attackers can break the resource isolation of Wasm runtimes. To achieve this, we analyze the most widely used Wasm runtimes to identify the accessible resources and related interfaces that Wasm instances can exploit, which are the attack surface of Wasm runtimes. In order to explore the attack surface systematically, we propose a static analysis framework, which can perform a comprehensive analysis of the selected Wasm runtimes. However, there are several challenges in the analyzing process. Firstly, existing code intermediate representation (IR) extraction tools do not adequately support the analysis of Rust projects, and both of the analyzed Wasm runtimes are Rust projects. Additionally, embedded assembly instructions in the project code lack proper representation, making it challenging to fully construct control flow and data flow dependencies. In addition, the analyzed project code heavily relies on Rust-specific features, resulting in a substantial number of indirect calls. This significantly complicates the resolution of call relationships and impacts the precision of control flow analysis. To address these challenges, we propose several new static analyzing methods. First, we propose a new Rust-oriented IR extraction approach to extract the complete IR of Rust projects, providing foundational support for the further analysis. Next, we design a refactoring-based IR generation approach to capture the IR of embedded assembly code, enhancing control flow and data flow information. Finally, we propose a Rust-specific indirect call resolution method, accurately resolving indirect call targets introduced by Rust-specific features and ultimately enabling precise control flow analysis. By using the proposed analysis methods, our analysis framework can identify potential attack surfaces systematically.

We categorize the identified attack surface and propose several exploitation strategies for each category. To validate the effectiveness of these strategies, we conduct experiments and assess their impacts. Based on the experiments, we demonstrate that attackers can exploit the WASI/WASIX interfaces to break the containerized environment created by Wasm runtimes, deplete system resources, and affect the host system and other Wasm instances. Finally, we discuss strategies to mitigate these attacks.

The main contributions of this paper are as follows:

\textbf{Attack Surface Measurement.} To measure the resource isolation attack surface of Wasm runtimes systematically, we propose a set of static analysis approaches that can address the analysis challenges introduced by the Rust-specific features of Wasm runtimes. To the best of our knowledge, this is the first static analysis approach for Rust-specific features, and we open source the prototype for further research. 

\textbf{Proposing Exploitation Strategies.} We propose several exploitation strategies targeting the identified attack surface of Wasm runtimes, focusing on the file system, I/O, network, etc. These strategies enable attackers to deplete system resources and influence the host system and other Wasm instances, which is a new kind of attack for Wasm containers.

\textbf{Evaluating the Impact.} We conduct experiments to evaluate the effectiveness of the proposed exploitation strategies and assess the security risks of the attack surface. Additionally, based on the experimental results, we discuss mitigation strategies.

The remainder of this paper is organized as follows: Section 2 provides the background on WebAssembly and discusses our observations. Our proposed attack surface analysis method is presented in Section 3. Section 4 introduces the exploit strategies we propose for the attack surface. Section 5 describes the experimental setup and evaluates the impact of these exploit strategies on system performance. The possible mitigation approaches for the exploit strategies are discussesed in Section 6. Section 7 gives the related work, and Section 8 concludes the paper.

\section{Background \& Observation}

To enable high-performance applications to run on the web, the W3C community developed the Wasm standard, a binary instruction format designed for stack-based virtual machines\cite{webassembly}. Developers can compile their source code into Wasm modules and run them in the browser. However, Wasm's design does not rely on web-specific features, which has led to its adoption in environments beyond the browser. For instance, there are standalone runtimes available that can run Wasm modules in server environments and allow developers to run Wasm modules within their own code. Due to its lightweight and cross-platform characteristics, Wasm is increasingly seen as a potential alternative to Docker\cite{gackstatter2022pushing,kjorveziroski2023webassembly,pham2023webassembly,stolz2023galois}.

\begin{figure}
    \centering
    \includegraphics[width=0.45\textwidth]{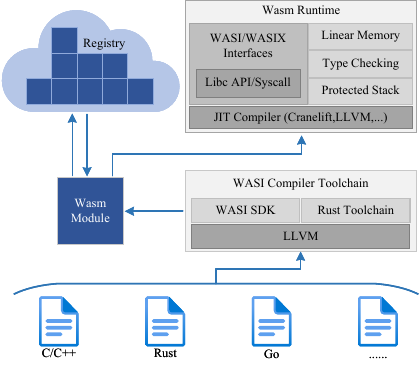}
    \caption{Overview of the Wasm Ecosystem. Code written in various languages can be compiled into Wasm modules using specific toolchains: the WASI SDK for C/C++ and the Rust toolchain for Rust. The Wasm modules can be packaged and uploaded to a registry (wasmer.io), from which they can be downloaded when needed. The Wasm runtime executes the compiled Wasm modules and provides the WASI/WASIX interfaces for them to access the underlying operating system. Additionally, Wasmtime includes several security mechanisms, such as linear memory and type checking, to ensure safety.}
    \label{fig:wasm-runtime-usage}
\end{figure}

Wasm and Docker containers share similarities, such as being lightweight and supporting cross-platform environments, but they exhibit significant differences in several areas. First, they differ in their packaging methods. Docker packages both the target application and its required environment into a standalone container image, typically without modifying the target application itself. In contrast, Wasm compiles the program's source code into Wasm modules, which may require certain modifications to the program. Second, their isolation mechanisms also differ. Docker leverages Linux features, such as Namespaces and Cgroups, to create an isolated environment when launching a container instance, thereby limiting the execution of the application within the container. In contrast, Wasm modules are run through a runtime, which is responsible for providing isolation functionality. Additionally, Wasm runtimes offer security mechanisms that restrict the memory usage of Wasm modules, mitigating the impact of various attacks (such as control flow hijacking or arbitrary memory read/write), which significantly enhances security.

The execution of Wasm instances depends on the Wasm runtime and the WASI/WASIX interfaces it provides. As shown in Figure \ref{fig:wasm-runtime-usage}, using the WASI compiler toolchain, developers can compile source code written in various languages into Wasm modules, which contain Wasm instructions. These Wasm modules can then be executed by the Wasm runtime. However, Wasm instructions themselves cannot directly interact with the underlying operating system (lack syscall). The underlying operating system services can only be used by invoking WASI/WASIX interfaces, which are provided by the Wasm runtimes. These interfaces also restrict the behaviors of Wasm instances by defining how Wasm instances interact with the operating system and specify which resources can be accessed. These interfaces function similarly to Seccomp with a whitelist, allowing Wasm instances to access only a subset of the operating system's functionalities (similar to limiting the syscalls available to Docker containers).

Currently, the most widely used Wasm runtimes under active development provide underlying system services through the WASI/WASIX interfaces. For instance, Wasmtime\cite{Wasmtime} is a Wasm runtime that supports the WASI interfaces. It enables the execution of Wasm modules outside of the browser while enforcing strict isolation policies. Wasmtime offers a command-line tool that allows Wasm modules to be run directly in the shell. Additionally, Wasmtime supports embedding itself into several programming languages, including Rust, C/C++, and Python. There is also a runtime called Wasmer\cite{Wasmer}, which supports the WASIX interfaces. Similarly, Wasmer provides a command-line tool and language-embedding capabilities for running Wasm modules. In fact, Wasmer functions as a WebAssembly ecosystem, similar to Docker. As shown in Figure \ref{fig:wasm-runtime-usage}, developers can use Wasmer to package Wasm modules and upload them to a registry, from where they can later download and execute the modules as needed. Wasmtime and Wasmer are currently the two most widely used Wasm runtimes, each with over 15K stars on GitHub. Both runtimes incorporate strict validation logic when implementing the WASI/WASIX interfaces to ensure security and isolation, such as enforcing filesystem isolation through path checks. The constraints imposed by the WASI/WASIX interfaces, combined with security mechanisms such as linear memory and type checking, along with the near-native execution speed of Wasm, make these runtimes secure and fast. 

However, we find that the resource isolation in these Wasm runtimes is not sufficiently robust, and in some aspects, it is even weaker than that of Docker containers. Specifically, we observe that these runtimes lack effective mechanisms to limit the consumption of system resources by Wasm instances. Since Wasm instances interact with the underlying operating system through the WASI/WASIX interfaces, these interfaces represent an attack surface exposed to Wasm instances. Attackers can exploit these interfaces to maliciously consume host resources, affecting the performance of both the host and other running Wasm instances. For example, a malicious Wasm instance can exploit file system interfaces to degrade system I/O performance to near zero by grabbing read/write bandwidth. These security problems are caused by the insufficient resource restriction mechanisms on the one hand, and on the other hand, the attacker can exploit the underlying mechanisms of the operating system through the WASI/WASIX interfaces by introducing workloads to other system components, so the protection is very difficult. As Wasm usage expands from browsers to servers, IoT, and cloud-edge computing, this attack surface may introduce various security risks. To fully understand these risks, a systematic exploration of the new attack surfaces introduced by WASI/WASIX is needed.

\section{Attack Surface Analysis}

\begin{figure*}
    \centering
    \includegraphics[width=0.95\textwidth]{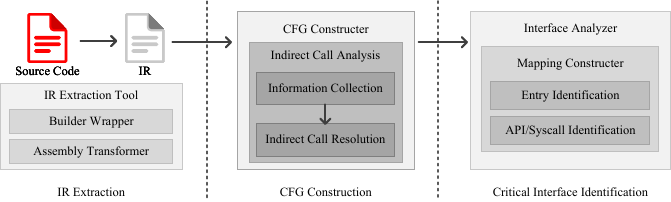}
    \caption{Overview of our analysis framework. The analysis framework comprises three stages: IR extraction, CFG construction, and critical interface identification. First, both source code and assembly are translated into IR. Next, the IR is used to build the control flow graph (CFG). Finally, WASI/WASIX interfaces are identified, and their mappings to underlying APIs or syscalls are established.}
    \label{fig:framework}
\end{figure*}

To systematically explore the attack surfaces due to insufficient restrictions, we delve into the exploitable points by analyzing the implementations of WASI/WASIX interfaces in Wasmtime and Wasmer. As the two most widely used Wasm standalone runtimes on GitHub, their implementations of the WASI/WASIX interfaces are representative. We start with static analysis of the source code to examine the specific implementations of the WASI/WASIX interfaces, collecting external APIs and underlying syscalls these interfaces depend on. Based on this information, we identify vulnerable WASI/WASIX interfaces.

There are several challenges that complicate the analysis process. First, it is challenging to extract IR from Rust projects. For static analysis, it is important to extract a complete IR of the source code. The analyzed Wasm runtimes are implemented in Rust, but the current IR extraction tools (e.g., wllvm\cite{wllvm}) do not support extracting IR for Rust projects. Second, the presence of assembly code blocks in the IR affects control flow analysis. The Rust projects to be analyzed are complex and contain source files in other languages, such as C and assembly. When extracting the project's IR, high-level language constructs (such as functions and global variables) are translated into their corresponding IR representations. However, assembly code is not similarly transformed, and control flow cannot be constructed for these parts, which leads to incomplete analysis results. Finally, extensive indirect calls reduce the accuracy of the analysis. In analyzing the Wasm runtime code, we found that it relies heavily on Rust-specific features, such as asynchronous mechanisms and dynamic dispatch. These features introduce a large number of indirect calls into the IR, which makes it difficult to construct accurate control flow graph (CFG) in static analysis\cite{lu2019does,liu2024improving,xia2024deeptype,cai2024unleashing}.

To comprehensively explore potential issues within the WASI/WASIX interfaces, we propose a systematic static analysis framework designed to address the challenges encountered during the analysis process. To tackle the challenges, the framework incorporates several critical techniques, including a custom IR extraction tool tailored for Rust projects, a semantic parsing method for embedded assembly code, and a context-driven indirect call analysis strategy specifically designed for Rust. As illustrated in Figure \ref{fig:framework}, our proposed analysis framework comprises three main stages: the IR extraction stage, the CFG construction stage, and the critical interface identification stage. In the IR extraction stage, the framework customizes the project's build process to ensure that a complete IR corresponding to the source code is generated during compilation. Additionally, this stage converts assembly code blocks within the generated IR into equivalent IR representations, enabling unified behavior analysis across the entire project. During the CFG construction stage, the framework leverages the complete IR to construct a precise CFG. It also analyzes indirect calls, building comprehensive and accurate call relationships. Finally, in the critical interface identification stage, the framework identifies entry points of the WASI/WASIX interfaces and involved critical invocations from the CFG. By following this systematic analysis workflow, the framework can effectively explore the attack surface of WASI/WASIX interfaces. The key technologies are described as follows.

\subsection{Rust-oriented IR Extraction}

The extraction of IR is a critical step for static analysis, and its completeness and accuracy directly influence the effectiveness of analysis. However, the current compiling tools cannot generate complete IR for Rust projects.

To address this problem, we propose a Rust IR generation approach to automatically extract the complete project-level IR during the building process of Rust projects and implement it as a tool. Specifically, the tool customizes the build process by applying link-time optimization techniques. This approach forces the compiler to output bytecode during the compilation, rather than machine instructions. Since the runtime source code includes both Rust and C components, we adjust the compilation options for both rustc and clang. During the construction of the final binary, all bytecode is merged and embedded within it. In addition, to maintain semantic consistency between the bytecode and the source code, optimization is disabled to prevent unnecessary modifications. At the end, the tool extracts the bytecode from the compiled runtime binary and outputs an IR file corresponding to the runtime source code, which contains the complete semantic information of the entire project. Our tool operates without requiring modifications to the project source code, thereby avoiding the need to address symbol conflicts and inter-module dependencies, and ensuring accurate and complete IR input for subsequent analysis.

\subsection{Assembly Code Transformation}

After extracting the project's IR, it is necessary to further process embedded assembly code blocks. Functions defined in assembly are embedded as raw assembly blocks within the IR, rendering their semantics inaccessible to unified analysis. These unstructured code blocks disrupt the completeness of control flow construction and hinder subsequent static analysis. To address this issue, we take a further step to transform these assembly code blocks into representations consistent with the IR, thereby simplifying the subsequent analysis process. Specifically, our framework first compiles the assembly blocks into binary object files and then uses decompilation tools to convert them into high-level language representations. During this process, the generated code undergoes necessary adjustments, such as replacing tool-specific constructs, to ensure compatibility. Finally, the adjusted code is recompiled to produce standardized LLVM IR, which is then integrated into the global IR. This approach effectively addresses the semantic gaps caused by assembly code, laying a solid foundation for constructing accurate control flow graphs.

\subsection{Rust-specific Indirect Call Analysis}

Comprehensive and accurate collection of code context information is a critical step in code analysis. This process typically involves constructing CFGs, extracting function call relationships, and identifying key data dependencies based on the IR. However, indirect calls pose significant challenges to the accuracy of this information. If the target analysis of indirect calls is imprecise, it will lead to inaccuracies in the collected CFGs, call relationships, and data dependencies, ultimately compromising the reliability of the analysis results.

During the analysis of Wasm runtimes, we observed that the IR bitcode contains a large number of indirect calls. The implementation of Wasm runtimes heavily relies on Rust language features, such as dynamic dispatch and asynchronous mechanisms, which introduce a large number of indirect calls in the IR, complicating the construction of precise call relationships and impacting data dependency analysis. Currently, the predominant approach for resolving indirect calls is type-based analysis\cite{lu2019does}, which narrows the scope of possible indirect call targets by leveraging function signatures and type constraints. This method offers low computational complexity and is particularly suited for large-scale projects, reducing the number of potential targets for indirect calls. However, in Rust projects with extensive use of Rust language features, function signatures and usage patterns often exhibit high similarity. This diminishes the effectiveness of type-based methods, making it challenging to achieve sufficiently precise results.

List \ref{lst:rust_indirect_call} shows a typical example illustrating how various Rust mechanisms result in indirect calls. In the source code of the runtime, situations similar to line 6-10 in List \ref{lst:rust_indirect_call} often occur, where an object (dir\_entry) is retrieved from a hard-to-trace memory object (table), and then a method (open\_file) is called based on it. However, the table, when defined, leverages Rust's dynamic dispatch mechanism, specifying that it can store a trait object instead of a specific type. A trait defines a set of interfaces, and each type can have its own independent set of implementations. This approach results in indirect calls being generated at compile time, where the specific implementation of the trait is called based on the actual type of the retrieved object. Additionally, after calling the open method, there is a special call to .await, which involves another Rust mechanism: asynchronous functions. This mechanism also leads to indirect calls being generated during compilation. The IR generated for asynchronous functions differs significantly from the source code, where it is restructured into two separate functions. One function (IR\_path\_open) creates a wrapper containing the function's context information, while the other function (IR\_path\_open\_closure) implements the original function's functionality based on that wrapper. Furthermore, a separate poll method is generated to advance the execution of the original function, utilizing an indirect call. Both mechanisms shown in List \ref{lst:rust_indirect_call} lead to the generation of indirect calls in the IR, and the analysis of one indirect call may depend on the analysis of another, making the overall analysis more complex. The type-based method can identify a set that contains the real indirect call targets. However, the function signatures of these indirect calls are often very generic, leading to a large set, which severely impacts the accuracy of the analysis.

\begin{lstlisting}[
caption={Rust code snippets and corresponding IR.},
label={lst:rust_indirect_call},
language=llvm,
backgroundcolor=\color{white},
basicstyle=\ttfamily\footnotesize,
numbers=left,
numberstyle=\tiny\color{gray},
stepnumber=1,
numbersep=5pt,
showspaces=false,
showstringspaces=false,
showtabs=false,
frame=l,
linewidth=0.45\textwidth, 
xleftmargin=15pt,
tabsize=2,
captionpos=b,
breaklines=true,
breakatwhitespace=false,
escapeinside={\%*}{*)},
]
// original function in source code
async fn open_file(...) -> ... {}
async fn path_open(...) -> ... 
{
...
let dir_entry = table.get_dir(...)?;
...
let file = dir_entry.dir
            .open_file(...)
            .await?;
...
}
// corresponding IR
define @IR_open_file(...) {}
define @IR_open_file_closure(...) {}
define @IR_path_open(...) {}
define @IR_path_open_closure(...)
{
...
// %123 is function pointer of IR_open_file.
invoke %123(...)
...
invoke @poll(...)
...
}
define @poll(...)
{
...
// %234 is function pointer of IR_open_file_closure.
invoke %234(...)
...
}
\end{lstlisting}

To achieve more precise indirect call analysis, we propose a specialized approach for analyzing indirect calls in Rust. Building upon the type-based methods, our approach further narrows the potential set of indirect call targets. This method takes full advantage of the semantic patterns of Rust language features in the LLVM IR, leveraging the deep insights from the patternized code generated at compile time to refine the indirect call analysis process. Our approach is composed of two steps: information collection and indirect call resolution, as shown in Figure \ref{fig:framework}. In the first step, we identify and record vtables. We observe that when trait objects are stored in the IR, both the object data and the corresponding trait's vtable are stored together. The vtable stores function pointers, which are then used to invoke methods with the associated object data. In the IR, we identify vtables by analyzing the properties of global objects, and then record the vtable along with the corresponding trait. In the second step, we apply corresponding analysis methods based on the type of indirect calls. At each indirect call site, we first analyze the data flow to identify the source of the indirect call. If the indirect call is caused by dynamic dispatch, we match the source's type information with the recorded vtable to find the corresponding vtable. If the indirect call is caused by asynchronous mechanisms, we use data flow analysis to trace the source of the pointer. For other types of indirect calls, we continue to use type-based analysis methods. Furthermore, since the analysis of one indirect call may depend on the analysis of another one, our approach is iterative and continues until no new indirect calls can be resolved.

For the two indirect calls at lines 21 and 29 in List \ref{lst:rust_indirect_call}, the proposed indirect call analysis can precisely identify their targets. In the first step, our approach can identify the global variable representing the vtable for the WasiDir trait. By parsing the variable, we obtain the function pointers within it and their offsets. Then, in the second step, the sources of the function pointers used at lines 21 and 29 are analyzed. These calls are related to Rust's dynamic dispatch and async mechanisms, respectively. Specifically, the indirect call at line 21 is associated with the WasiDir trait. We use data-flow analysis to determine which function pointer in the corresponding vtable is used. The indirect call at line 29 is related to the async mechanism. We apply backward data-flow analysis to trace its function pointer. However, this function pointer is returned from the indirect call at line 21. Therefore, our analysis first analyzes the target of the call at line 21. This example shows that analyzing one indirect call may depend on another indirect call, so our approach can iteratively analyze indirect calls.

During our analysis of indirect calls, we find some false positives. After collecting and analyzing a number of these false positives, we find that the use of generic functions is a major cause of such cases (e.g., call\_unchecked\_raw and call\_raw in the Wasmtime source code). Generic functions serve as templates, and at each call site that invokes them, the compiler generates a specialized function instance based on the template. Since generic functions are not specific to Rust, we employ MLTA \cite{lu2019does,liu2024improving}, a state-of-the-art type-based indirect call analysis technique, to analyze these cases. When a generic function defines how to perform an indirect call, this behavior becomes fixed in the template. As a result, all specialized instances invoke the indirect call in the same manner, leading MLTA to consider that each call site may invoke the same set of call targets. However, in practice, each function instance may only call a subset of those targets. This discrepancy causes MLTA to over-approximate the set of possible targets, resulting in false positives. This is an inherent challenge in static analysis.

\subsection{Attack Surface Exploration}

To analyze how attackers can attack the underlying operating system through the WASI/WASIX interfaces, we propose a systematic analysis approach aimed at systematically analyzing the accessible syscalls or APIs of each WASI/WASIX interface.
After constructing the call graph, the WASI/WASIX interfaces are collected and used as entry points for the analysis. Starting from these entry points, all possible call chains are traversed to identify all potential syscalls or APIs that may be invoked. For each encountered syscall or API, a program slice is generated that describes how the execution reaches this point from the entry point. We then apply static taint analysis to determine whether the parameters of the interface can directly or indirectly influence the parameters of the syscall or API. For example, an argument of the interface function may be passed down and eventually used as a syscall argument. In addition, we perform backward data-flow analysis from the syscall or API call sites to trace the origins of their arguments. In some cases, syscalls are invoked through the syscall() function, and we need to identify the specific syscall number used. Besides, some syscall arguments are sensitive to security, so we try to trace their concrete values. After completing the above analysis, we are able to determine, for each WASI/WASIX interface, the possible syscalls and external APIs it may invoke through the call chain, as well as some relationships among their parameters. As a result, the results of Table \ref{tab:runtime_dep} are obtained. Our approach helps analysts quickly locate critical issues and reduces the need for manual code review. In practice, the analysis completes in 47 minutes and 33 seconds. While our current focus is on Wasm runtimes, the proposed approach is also applicable to other Rust projects, demonstrating broad applicability.

As shown in Table \ref{tab:runtime_dep}, we list part of the results from our analysis. The first column in the table lists the analyzed runtimes, while the second column details the syscalls and external APIs used in their WASI/WASIX interface implementations.
Guided by these findings, we infer the attack strategies for these resource exhaustion issues based on three key aspects. First, we find that the collected syscalls can be used to consume system resources, so a straightforward idea is to directly exhaust host resources. This can be done by exhausting I/O through readv/writev and read/write, saturating network bandwidth with sendto and send, or consuming large amounts of disk space, etc. In addition, prior works \cite{sharma2016containers,gao2019houdini,yang2021demons,xiao2023attacks} have investigated resource exhaustion attacks against host systems in virtual machine or container environments (e.g., Docker containers). Inspired by them, we design similar strategies for Wasm runtimes to verify whether they are still effective. Furthermore, we identify new attack surfaces of Wasm runtimes that have not been analyzed in prior work. To assess the potential security implications, we analyze the security threats that these new attack surfaces may cause and design the corresponding attack strategies to validate their effectiveness. For instance, our analysis result shows that the statx syscall is used to retrieve file metadata within the WASI/WASIX interface implementation. And our approach reveals that its parameters include the AT\_STATX\_SYNC\_AS\_STAT flag, which triggers data synchronization and negatively impacts host I/O performance. Based on this insight, we propose new exploit strategies, and the experiment confirms their effectiveness. Finally, since the Wasm instance can access all WASI/WASIX interfaces, we also designed attack strategies that combine multiple interfaces. For example, by combining multithreading interfaces with other attack methods, the impact of the attacks can be amplified, posing a greater threat to the host system.

\begin{table}
    \centering
    \begin{tabular}{cp{5cm}}
        \toprule
        Runtime & Dependence \\
        \midrule
 	Wasmtime & openat unlinkat fsync fdatasync readv writev preadv pwrite64  pthread\_create\\
        \midrule
        Wasmer & open64 unlink read write sendto send statx pthread\_create\\
        \bottomrule
    \end{tabular}
    \caption{Runtimes and the syscalls/APIs related to their WASI/WASIX interfaces.}
    \label{tab:runtime_dep}
\end{table}

\section{Exploiting Strategies}

Based on the static analysis results, potential attack strategies are explored, which can be categorized into two categories. The first category uses Wasm instances to directly consume system resources, such as running tasks with high computing or I/O workloads, aiming to exhaust the system's resources and thus affect other Wasm instances. The second category is to exploit some Wasm runtime or system mechanisms to inject workloads into some system or runtime components so that the system resources are also exhausted but are not observed to be consumed by the malicious Wasm instance. Consequently, the current resource control mechanisms, such as Cgroups, cannot protect against these attacks.

\subsection{Direct Resource Consumption}

As shown in Table \ref{tab:runtime_dep}, Wasm instances can access interfaces related to the file system, I/O, and networking, enabling them to directly consume these resources. Due to the lack of effective resource limitation mechanisms (e.g., Cgroups) in Wasm runtimes (e.g., Wasmer and Wasmtime), malicious Wasm instances can consume critical system resources (e.g., CPU, disk, and memory resources), significantly degrading the performance of other instances hosted on the same machine. 

One of the most easily exploited resources is CPU cycles. By executing compute-intensive tasks, a malicious Wasm instance can exhaust CPU resources, thereby severely impacting the performance of other instances on the host. To address this issue, Wasmtime introduces the fuel mechanism, which tracks the number of executed instructions and terminates the Wasm instance once a predefined threshold is reached. Wasmer offers similar resource control mechanisms. However, these mechanisms are unsuitable for long-running server-side applications, and forced termination may disrupt service continuity or lead to functional anomalies.

In addition to CPU cycles, malicious Wasm instances can consume I/O resources through disk accesses, such as storing large files. This attack is possible as long as the Wasm instance is allowed to access files or directories, and this scenario is very common. Furthermore, malicious Wasm instances can occupy I/O bandwidth by creating numerous files and continuously writing data into them, impacting other instances that are sensitive to I/O bandwidth. 

While memory resources can also be consumed by malicious Wasm instances, the linear memory model of Wasm enforces limitations on accessible memory spaces (typically 4GB). But this mechanism can only mitigate the security risk of a single instance. Previous research\cite{ristenpart2009hey} has found that attackers have the ability to run multiple malicious instances on the same physical machine. In this scenario, the resource limitation policy for a single instance cannot work. 

\subsection{Exploiting File System Interfaces}

Wasm instances can access file systems through WASI/WASIX interfaces provided by Wasm runtimes. While some security mechanisms are embedded within interface implementations to validate path access permissions and prevent unauthorized operations, these mechanisms fail to effectively restrict file system resource consumption. In addition to disk resource consumption, malicious Wasm instances can exhaust critical kernel resources (e.g., inode resources, system entropy pool, etc.) in the underlying operating system through the accessible file-related WASI/WASIX interfaces, some of which invoke resource-consumption syscalls of the underlying operating system. By creating a large number of files, a malicious Wasm instance can rapidly consume inode resources, making the operating system unable to create new files. This approach can seriously affect the file access performance of other instances, leading to serious instability of the whole system. Even if inode resources are not completely exhausted, the proliferation of inode nodes in the kernel can significantly degrade system performance, and frequent file creation operations can also degrade disk performance. Continuous reads from /dev/random can deplete the system entropy pool, blocking other processes that depend on this system service.

In addition, the underlying file access capabilities provided by the WASI/WASIX interface can allow a malicious Wasm instance to occupy some limited number of operating system resources, thus affecting the normal execution of other instances. For instance, the /dev/ptmx in the Linux virtual filesystem can be accessed by Wasm instances through the WASI/WASIX interfaces, which is used to get file descriptors for pseudo-terminals. However, the number of /dev/ptmx files that can be opened concurrently is limited. A malicious Wasm instance can easily exhaust this resource, leading to severe impacts on pseudo-terminal-related system services and potentially causing DoS attacks to related Wasm instances. 

Furthermore, file accesses can allow malicious Wasm instances to trigger other system events (e.g., context switching, etc.) that degrade system performance. When an instance performs I/O requests with high frequency, and even though each I/O operation involves only small amounts of data, the high-frequency invocation can result in context switching and high disk seek time, reducing the system's I/O performance. Additionally, malicious Wasm instances can inject workloads into the Wasm runtime. For instance, attackers can invoke WASI/WASIX interfaces with logging capabilities and make the Wasm runtime consume host I/O resources secretly by generating an excessive volume of log entries.

Beyond basic file system I/O interactions, Wasm instances can leverage WASI/WASIX interfaces to trigger file data synchronization in the underlying operating system, thereby inducing performance disruptions of the operating system. Specifically, two critical synchronization functions (i.e., fsync and fdatasync) can be invoked through WASI/WASIX interfaces, which can force metadata and content synchronization from cache to disk. These synchronization operations generate significant disk I/O overhead, consequently impacting system-wide disk performance. Since most of these attacks are capable of injecting workloads into other components of the underlying operating system, it is difficult to mitigate these attacks through resource constraints on Wasm instances.

\subsection{Exploiting Network Interfaces}

Wasm runtimes provide network interfaces that allow Wasm instances to establish connections and transmit data. However, the lack of network resource control mechanisms exposes systems to potential resource attacks. Malicious Wasm instances can exploit this deficiency through two strategies. The first strategy involves directly consuming network bandwidth. Since Wasm runtimes typically do not enforce bandwidth management policies, malicious instances can continuously transmit large volumes of data, monopolizing network bandwidth. This behavior disrupts the operation of other instances that rely on network resources, leading to performance degradation. The second strategy employs network interfaces to inject workloads to the underlying operating system. The operating system needs to utilize mechanisms and components such as interrupts, kernel threads, etc., to process network packets. Even though small data packets consume little bandwidth, they can introduce kernel processing overhead. Therefore, malicious Wasm instances can perform frequent small-packet transmissions to significantly increase CPU load and impact the system's overall network performance.
As these attacks can inject workloads into other parts of the underlying operating system, it's difficult to defend against such attacks by imposing resource limitations on Wasm instances.

\subsection{Exploiting Thread Mechanisms}

The multithreading interfaces provided by Wasm runtimes enable Wasm instances to magnify their resource usage and attack effectiveness through concurrent operations. Specifically, multithreading amplifies the impact of previously discussed exploitation strategies in three aspects. First, multithreading can accelerate the consumption of system resources. By executing tasks concurrently, malicious Wasm instances can exhaust critical resources, such as CPU, memory, and disk, at an accelerated rate, facilitating attack effects. In addition, multithreading provides resource-grabbing advantages for attackers. Malicious instances can perform concurrent operations to outperform single-threaded competitors in acquiring system resources. Furthermore, the scheduling and management of a large number of threads can also introduce additional pressure on the underlying operating system kernel. 

\section{Experiments}

In this section, we evaluate the effectiveness of the proposed exploitation strategies. After setting up the experimental environment, we analyze the impact of exploiting the two Wasm runtimes (i.e., Wasmtime and Wasmer) under the proposed exploitation strategies, respectively, and discuss the experimental results.

\subsection{Experiment Setup}

We conduct the experiments in a controlled local environment to minimize the potential impact. A local area network is established and isolated from the external environment, ensuring that our experiments do not affect real-world operations. The experimental setup is based on a machine equipped with an Intel Core i7-10700 CPU, 16 GB of RAM, a 1 TB HDD, and a Gigabit Ethernet network interface card. The operating system used is Ubuntu 24 with kernel version 6.8.0-51-generic. The versions of the Wasm runtimes analyzed in this study are Wasmtime 24.0.0 and Wasmer 4.3.2.

To assess the impact of the proposed strategies, we focus on three primary metrics: CPU usage, file I/O bandwidth, and network bandwidth. During the experiments, we use pidstat\cite{pidstat} to measure the CPU usage of the malicious Wasm instance, dstat\cite{dstat} to monitor overall system CPU idle time, fio\cite{fio} to evaluate changes in I/O performance, and iperf3\cite{iperf3} to assess network performance fluctuations. Additionally, we use ent\cite{ent} to test the effects of exhausting the entropy pool. Prior to the experiments, we minimize the interference of background and foreground processes by terminating all non-essential tasks. We then measure the system performance to establish a baseline for comparison. According to the dstat results, the idle CPU usage is 99.66\%. The fio benchmark shows a file read performance of 456.36 MB/s and a write performance of 237.50 MB/s. The iperf3 results indicate network upload and download bandwidths of 874.90 Mbits/s and 872.54 Mbits/s, respectively. Finally, the random number generation rate from /dev/random is measured at 94.20 MB/s.

\subsection{Experiments on Wasmtime}

For Wasmtime, its file system and I/O interfaces can be exploited to consume the host machine's resources. However, since Wasmtime currently does not support active network connections initiated by Wasm instances, exploiting this point is not practical. Therefore, we do not conduct experiments involving network exploitation. The experimental results are presented in Table \ref{tab:wasmtime_exp}.

\begin{table}[h]
    \centering
    \begin{tabular}{ccccc}
        \toprule
        operation & cpu\_pid & cpu\_idl & fio\_r & fio\_w \\
        \midrule
        baseline & - & 99.66\% & 456.36 & 237.50 \\
        \midrule
        sync data & 10.14\% & 75.13\% & 28.09 & 0.10 \\
        sync all & 10.08\% & 75.09\% & 27.76 & 0.10 \\
        file op & 4.45\% & 84.54\% & 6.48 & 4.08 \\
        \midrule
        write(S) & 2.31\% & 87.85\% & 2.81 & 2.92 \\
        write(L) & 0.88\% & 85.88\% & 1.34 & 0.68 \\
        read(S) & 99.89\% & 0.00\% & 96.03 & 54.22 \\
        read(L) & 0.09\% & 58.41\% & 0.04 & 0.02 \\
        \midrule
        W null & 99.70\% & 0.00\% & 54.76 & 22.58 \\
        W tty & 5.59\% & 87.24\% & 421.97 & 216.52 \\
        R zero & 99.76\% & 0.00\% & 57.84 & 28.85 \\
        R random & 99.92\% & 0.00\% & 85.22 & 47.75 \\
        \bottomrule
    \end{tabular}
    \caption{Impact of exploiting Wasmtime's interfaces on system performance. The first column lists the types of operations. The second column shows the CPU usage of the process running the Wasm instance, while the third column reports the overall system idle CPU usage during the test. The fourth and fifth columns indicate file read and write bandwidths (in MB/s). The first row represents baseline performance under idle conditions. Rows 2 and 3 correspond to data synchronization-related operations, respectively. Row 4 presents file creation and deletion operations. Rows 5–8 detail file read/write operations, with \textbf{S} denoting small chunks and \textbf{L} representing large chunks. The final four rows summarize read or write operations on device files, with \textbf{W} denoting write and \textbf{R} representing read.}
    \label{tab:wasmtime_exp}
\end{table}

\textbf{Exploiting file operation interfaces.} When implementing the WASI interfaces, Wasmtime leverages the cap\_std crate (a Rust standard library), utilizing its functionality to support various WASI operations, such as data synchronization and file handling. By examining the call chains of data synchronization-related interfaces, we find that the libc functions fsync and fdatasync are ultimately invoked. As previously discussed, these operations can severely impact disk I/O performance. To quantify the effect, we initiate a Wasm instance to continuously invoke these interfaces, and record their impact on the operating system's performance. The results are presented in rows 2-3 of Table \ref{tab:wasmtime_exp}. The data show that both types of operations lead to a reduction in file read and write performance by more than 93\% and 99\%, respectively, compared to the baseline. Furthermore, according to the data in the table, the CPU usage of the Wasm instance is approximately 10\%, while the overall system CPU usage is around 25\%, indicating an additional computational overhead of 15\%. This suggests that, during the execution of data synchronization tasks, some of the workload is not accounted for by the current process, making it difficult to restrict these attacks using the current Linux resource isolation mechanisms (e.g., Cgroups).

For file handling interfaces, by examining the call chains of file creation and removal operations, we find that the syscalls openat2 and unlinkat are triggered, through inline assembly. Frequent invocation of these interfaces can also have a significant impact on system performance. After having a Wasm instance repeatedly invoke these interfaces, we record how the system performance changes, as shown in row 4 of Table \ref{tab:wasmtime_exp}. Compared to the baseline, system I/O performance decreases by over 99\%. Notably, the CPU usage of the Wasm instance remained at only 4.45\%, while the system's idle CPU usage was 84.54\%, indicating an additional overhead of approximately 10\%. Similar to data synchronization operations, the performance degradation caused by file handling operations cannot be fully mitigated by the Cgroups mechanism.

\textbf{Exploiting file R/W interfaces.} Wasmtime provides four I/O interfaces, two of which offer traditional file read/write functionality, while the other two support reading/writing at specified offsets within a file. By examining the call chains of these interfaces, we find that they finally invoke the readv/writev functions and the preadv/pwritev syscalls. Without proper restrictions, a single Wasm instance can easily exhaust I/O resources, leading to a significant degradation in the system's overall file read/write performance. To quantify the impact on system performance, we continuously perform file read/write operations on disk within a Wasm instance. Furthermore, we vary the chunk sizes used in these operations to analyze performance under different I/O patterns. The results are summarized in rows 5–8 of Table \ref{tab:wasmtime_exp}. According to the data presented in the table, operations involving large chunks lead to significant degradation in read and write performance (over 99\%), while causing minimal CPU usage, measured at 0.09\% and 0.88\% for read and write operations, respectively. Similarly, writing small chunks also results in severe performance degradation, albeit with slightly higher CPU usage at 2.31\%. In contrast, reading small chunks exhibits different characteristics: the Wasm instance incurs a CPU usage of 99.89\%, but the corresponding impact on system I/O performance is relatively modest, at approximately 77\%. Based on the results, we can find that these operations can induce additional CPU overhead to other components of the operating system, so limiting the I/O usage of Wasm instances cannot prevent Wasm instances from consuming other system resources, thus affecting other instances.

If Wasmtime is configured to allow access to the /dev directory, a Wasm instance can read and write device files within it. The last four rows of Tables \ref{tab:wasmtime_exp} illustrate the impact of performing read and write operations on these files, with \textbf{W} denoting write and \textbf{R} representing read. Specifically, these rows detail the effects of reading from or writing to four files: /dev/null, /dev/tty, /dev/zero, and /dev/random. As shown in Table \ref{tab:wasmtime_exp}, when writing to /dev/null and reading from /dev/zero, the Wasm instance exhibits excessive CPU usage, leading to a significant degradation in I/O performance (over 87\%). The impact of these exploiting strategies can increase the workloads of the malicious Wasm instances can consume the host resources. 

In contrast, accessing the virtual devices in /dev can introduce workloads to other system components of the underlying operating system. Reading from /dev/random can deplete the entropy pool, causing blocking for other processes dependent on it. According to the experimental results, when the Wasm instance continuously reads from /dev/random, the random number generation rate, as measured by the ent test, is 7.60 MB/s, representing a 91.93\% decrease compared to the baseline, which can be used to perform the DoS attack and is difficult to defend against. In addition, writing to /dev/tty can introduce additional CPU usage that cannot be effectively limited using Cgroups. 

Additionally, within the /dev directory, there is a file called /dev/ptmx, which is used to create master-slave pairs for pseudo-terminals, commonly employed by applications like xterm and sshd. The number of open instances of this file is subject to a low limit, which can be viewed in /proc/sys/kernel/pty/max, typically set to 4096. In our experimental environment, this limit is significantly lower than the maximum number of file descriptors that a process can open (1M, as viewed through ulimit). Therefore, if Wasmtime permits a Wasm instance to access the /dev directory, the Wasm instance can exhaust the available file descriptors for /dev/ptmx. Once this resource is exhausted, new shells cannot be opened, and SSH connections to the host machine become impossible. This vulnerability could be exploited by an attacker to launch a DoS attack, disrupting not only user access but also processes that rely on pseudo-terminals, severely compromising the availability of system services.

\subsection{Experiments on Wasmer}

Besides files I/O interfaces, the Wasmer allows Wasm instances to access the host's network, so the network interface can also be exploited. We evaluate the attack impacts on the operating system when exploiting these interfaces provided by Wasmer and highlight the differences between Wasmer and Wasmtime.

\textbf{Exploiting file operation interfaces.} Wasmer also provides data synchronization interfaces. Unlike Wasmtime, Wasmer heavily relies on the functionalities provided by Tokio\cite{tokio}, an asynchronous Rust runtime designed for implementing asynchronous mechanisms in Rust. In Wasmer's implementation, data synchronization-related interfaces are built on Tokio's poll\_flush. Analyzing the call chains of these interfaces reveals that, unlike Wasmtime, they do not invoke libc functions such as fdatasync or fsync. Instead, they rely on asynchronous mechanisms to wait for the completion of preceding write tasks. We exploit these two interfaces by launching Wasm instances via Wasmer, with the respective results presented in the second and third rows of Table \ref{tab:wasmer_exp}. 

According to the results, exploiting data synchronization-related interfaces leads to a decrease in system I/O bandwidth. The sync data operation primarily increases its own workload, consuming CPU resources and thereby degrading overall system performance. In contrast, the sync all operation exhibits a lower CPU utilization but introduces approximately 8\% CPU load to other system components, resulting in a more significant impact on I/O bandwidth. This discrepancy arises because the sync all operation additionally updates file metadata stored in memory, such as file size. During metadata updates, the statx syscall is frequently invoked with the AT\_STATX\_SYNC\_AS\_STAT flag, which triggers metadata synchronization operations and intensifies the impact on I/O performance.

\begin{table}[h]
    \centering
    \begin{tabular}{ccccc}
        \toprule
        operation & cpu\_pid & cpu\_idl & fio\_r & fio\_w \\
        \midrule
        baseline & - & 99.66\% & 456.36 & 237.50 \\
        \midrule
        sync data & 99.96\% & 0.00\% & 178.24 & 62.85 \\
 	sync all & 5.75\% & 86.29\% & 7.73 & 3.82 \\
        file op & 5.00\% & 83.60\% & 8.03 & 4.59 \\
        \midrule
        write(S) & 8.17\% & 82.96\% & 2.60 & 2.76 \\
        write(L) & 0.98\% & 85.88\% & 1.42 & 0.68 \\
        read(S) & 71.64\% & 16.53\% & 124.15 & 59.91 \\
        read(L) & 0.11\% & 59.24\% & 0.06 & 0.04 \\
        \midrule
        W null & 96.57\% & 3.05\% & 208.06 & 99.03 \\
        W tty & 94.85\% & 0.24\% & 190.15 & 104.94 \\
        R $*$zero & 12.86\% & 85.56\% & 290.59 & 185.67 \\
        R $*$random & 5.93\% & 87.91\% & 429.76 & 219.96 \\
        W $*$null & 13.83\% & 84.45\% & 313.29 & 164.05 \\
        \bottomrule
    \end{tabular}
    \caption{Impact of exploiting Wasmer's file-related interfaces on system performance. The first column lists the types of operations. The second column shows the CPU usage of the process running the Wasm instance, while the third column reports the overall system idle CPU usage during the test. The fourth and fifth columns indicate file read and write bandwidths (in MB/s). The first row represents baseline performance under idle conditions. Rows 2 and 3 correspond to data synchronization-related operations, respectively. Row 4 presents file creation and deletion operations. Rows 5–8 detail file read/write operations, with \textbf{S} denoting small chunks and \textbf{L} representing large chunks. The final six rows summarize read or write operations on device files, with \textbf{W} denoting write and \textbf{R} representing read. Devices with an \textbf{$*$} prefix indicate that they were opened from a mapped directory, while those without the prefix were opened directly.}
    \label{tab:wasmer_exp}
\end{table}

In the case of Wasmer's file handling interfaces, an analysis of the call chains for file creation and deletion operations reveals that the open and unlink functions from libc are invoked. Similar to the situation with Wasmtime, frequent invocations of these interfaces can severely degrade the overall system performance. To evaluate this impact, we conduct an experiment by running Wasm instance to continuously invoke these interfaces, with the results shown in row 4 of Table \ref{tab:wasmer_exp}. The Wasm instance consumes only 5.00\% of the CPU but causes a more than 98\% reduction in I/O bandwidth compared to the baseline. Additionally, the Wasm instance injects approximately 11\% extra CPU load into the underlying operating system. Since file processing operations do not directly consume I/O bandwidth, the current resource limitation mechanisms in Linux are insufficient to mitigate the impact of this exploitation.

\textbf{Exploiting file R/W interfaces.} For the I/O interfaces provided by Wasmer, by analyzing the call chains, we find that they finally invoke the read and write functions from libc. We run a Wasm instance using Wasmer to evaluate these interfaces, and the experimental results are presented in rows 5-8 of Table \ref{tab:wasmer_exp}. All four operations result in a degradation of system I/O performance. Large chunk read/write operations consume significant I/O bandwidth, causing a reduction of over 99\% in system I/O bandwidth, while their CPU utilization remains below 1\%. In contrast, small chunk read/write operations do not occupy substantial I/O bandwidth but still lead to a significant drop in system I/O bandwidth. This is due to frequent I/O operations triggering continuous context switches and interrupts, which ultimately degrade overall I/O performance. In addition, these operations introduce additional CPU load to other system components. Notably, large chunk read operations generate approximately 40\% of extra CPU overhead. This indicates that limiting the CPU usage of Wasm instances using Cgroups is insufficient to mitigate the impact of these operations.

Wasmer prohibits setting /dev as an accessible directory, and device files directly opened by Wasm instances are not actual device files. However, exploitation is still achievable. Wasmer allows mapping host directories to guest environments, enabling us to map /dev to an alternative directory, such as /dev1. This allows Wasm instances to access real device files through the /dev1 directory. We evaluate the exploitation of device files through Wasmer, and the results are summarized in Table \ref{tab:wasmer_exp}. Except for reading from /dev1/random, all other operations significantly degrade system I/O performance. Furthermore, accessing device files through /dev incurs substantial CPU overhead. In contrast, accessing device files via /dev1 results in relatively lower CPU overhead, differing from the behavior observed in Wasmtime. This discrepancy arises from Wasmer's implementation, which does not create a large number of file descriptors for device files accessible to Wasm instances. Nevertheless, Wasm instances still impact system I/O performance. Additionally, reading from /dev1/random has a distinct effect: it reduces the random number generation rate to 86.90 MB/s, a decrease of 7.75\% compared to the baseline.

\textbf{Exploiting network interfaces.} In Wasmer, the Wasm instance can establish network connections. There are two interfaces for sending network packets, namely sock\_send and sock\_sendto. Based on the call chains of these interfaces, the libc functions send and sendto are ultimately invoked. We test the exploitation on these interfaces by running a Wasm instance within Wasmer, and the results are presented in Table \ref{tab:wasmer_net}. Network bandwidth is reduced by all the evaluated operations, with sending large packets having a more pronounced impact. Sending large packets consumes a significant portion of the network bandwidth, thereby limiting the resources available to other instances. In contrast, sending small packets does not consume as much bandwidth, but frequent context switches and interrupts impose excessive load on the kernel, ultimately degrading overall system performance. Except for sending small packets over TCP, all other operations inject additional CPU load into the underlying system. Specifically, using the UDP protocol introduces approximately 12\% extra CPU overhead. As a result, even with Cgroup limitations on network bandwidth and CPU usage, the impact of exploiting network-related interfaces remains challenging to mitigate.

\begin{table}[h]
    \centering
    \begin{tabular}{ccccc}
        \toprule
        operation & cpu\_pid & cpu\_idl & ipf\_s & ipf\_r \\
        \midrule
        baseline & - & 99.66\% & 874.90 & 872.54 \\
        \midrule
 	tcp pkt(L) & 2.72\% & 89.96\% & 61.79 & 60.99 \\
        tcp pkt(S) & 98.96\% & 0.53\% & 865.00 & 862.74 \\
        udp pkt(L) & 79.82\% & 8.22\% & 605.23 & 604.06 \\
        udp pkt(S) & 80.23\% & 8.17\% & 804.68 & 803.01 \\
        \bottomrule
    \end{tabular}
    \caption{Impact of exploiting Wasmer's network-related interfaces on system performance. The first column represents the operation type, where \textbf{L} denotes large packets and \textbf{S} denotes small packets. The second column shows the CPU usage of the Wasm instance. The third column indicates the system idle CPU usage during testing. The fourth and fifth columns report the bandwidth (in Mbits/s) for sending and receiving packets, respectively. The first row corresponds to the baseline performance when the system is idle.}
    \label{tab:wasmer_net}
\end{table}

\section{Security Enhancement Discussion}

Malicious Wasm instances can consume system resources, leading to severe performance degradation. And, many attacks can inject workloads into other components of the system, thereby affecting overall system performance. Such attacks are relatively covert and challenging to defend against. Here we discuss potential defense mechanisms.

\textbf{XFS}: The XFS file system provides management features that can limit the consumption of disk space and inode numbers. However, the current Wasm runtimes do not natively support XFS. To limit a Wasm instance, an XFS file system can be created with restrictions on storage space and inode count. Subsequently, the Wasm runtime can be configured to allow the Wasm instance to access only specific directories within this file system. If the Wasm instance exceeds the allocated disk space or inodes, XFS can block the corresponding operations, effectively preventing resource exhaustion.

\textbf{Cgroups}: Cgroups can be used to limit resource consumption such as CPU, pids, disk I/O, and network bandwidth available to a process. The current Wasm runtimes do not surport Cgroups configuration. To limit a Wasm instance, we can create and assign a process to a Cgroup prior to executing the Wasm runtime, thereby ensuring that the running Wasm instance is subject to the resource constraints imposed by the Cgroup. By limiting the cpuset, CPU time quotas, and the number of processes/threads, excessive CPU resource consumption by the Wasm instance can be prevented. Limiting I/O bandwidth and the frequency of I/O operations can prevent the Wasm instance from over-consuming I/O resources, thereby alleviating its impact on overall system I/O performance. Similarly, limiting network bandwidth can help reduce the Wasm instance's consumption of network resources.

Although the aforementioned measures can limit direct resource exhaustion by Wasm instances, malicious instances can still bypass them. For example, even with cgroups restricting CPU, threads, and bandwidth, abusing file interfaces can degrade I/O performance. In Wasmtime, data synchronization interfaces can also severely impact I/O without high CPU or bandwidth usage. Since Wasm instances share the same OS, they can interfere with each other through it.
In addition to being exploitable for DoS attacks, WASI/WASIX interfaces may also be used by attackers to achieve other goals. Prior works\cite{jiang2024sync+,chen2024write+} have shown that data synchronization mechanisms can be leveraged to build covert channels. For Wasm runtimes, it is also possible to achieve similar effects by abusing sync-related interfaces.

To further mitigate the impact of the proposed attacks, more fine-grained restriction mechanisms are needed. Existing work on syscall interposition\cite{yasukata2023zpoline,jacobs2024system,schrammel2022jenny} allows intercepting or filtering syscalls either in user space or kernel space. However, since the syscalls used by WASI/WASIX interfaces are common and required by many legitimate applications, simply disabling them \cite{christou2023binwrap,abbadini2023cage4deno} is not a practical solution. A more effective approach may combine the syscall interposition with frequency-based or learning-based defenses. Specifically, system administrators can define acceptable usage frequencies for relevant syscalls in advance or apply deep learning techniques to learn normal usage patterns. These profiles can then be used to constrain application behaviors. If an application deviates from expected patterns, the system can generate alerts or terminate its execution. For example, if an application triggers sync-related syscalls too frequently, syscall interposition can be used to block further access to those syscalls. This restriction approach requires expert knowledge, as applications vary widely in types and behaviors, making it necessary to craft specific patterns for each case. While this approach lacks generality, it may still be sufficient for specific use cases. Our proposed attack strategies are effective primarily because the relevant resources are not properly managed. Therefore, it is necessary to accurately detect and prevent processes from exhausting resources. eBPF is a suitable technology for this purpose, as it enables efficient monitoring without requiring kernel code modifications. \cite{shen2024towards} proposes a more fine-grained defense mechanism that uses eBPF to monitor and limit a process's consumption of various kernel abstract resources. This allows for more precise control over resource usage. However, defending against our proposed attacks also requires precise analysis and identification of the actual resource consumers, which may require extensive modifications to the kernel. In addition, monitoring too many resources may introduce performance overhead to the operating system. A practical compromise is to focus only on monitoring and analyzing sensitive resources, which can reduce the impact of performance while still offering effective protection.

\section{Related Work}

Wasm's flexible cross-platform deployment capability and near-native execution speed have enabled its application beyond the browser, with many works exploring its use in IoT and cloud-edge computing\cite{ribeiro2024wasmico,kuribayashi2023dynamic,pham2023webassembly,sekigawa2023web,lim2023advanced,kakati2023webassembly}. Recently, \cite{cao2024stateful} proposes an authorization framework based on Wasm for the cloud, aiming to achieve stateful minimal-privilege authentication. Additionally, Wasm has played a role in the field of zero-knowledge proofs. \cite{gao2024zkwasm} presents a virtual machine, called ZKWASM, to ensure trustworthy computation in cloud, edge, and grid computing. ZKWASM can simulate the execution of WASM bytecode and utilize ZKSNARK technology to generate corresponding zero-knowledge proofs, which can be used to convince entities without revealing confidential information. \cite{wang2024ligetron} implements a sublinear zero-knowledge system, Ligetron, which can be deployed in the browser via Wasm. \cite{zhao2023reusable} designs a reusable enclave with Wasm as the backend executor, significantly reducing the cold start overhead of TEE. With the widespread adoption of Wasm, its security concerns continue to receive increasing attention. Currently, research on Wasm security primarily focuses on vulnerabilities within Wasm, new attacks based on Wasm, and the development of novel security mechanisms. Additionally, some studies investigate other aspects of Wasm, such as performance, behavioral analysis, and readability.

\textbf{Vulnerabilities \& Attacks.} Some researches\cite{romano2021empirical,lehmann2020everything,zhou2023wadiff,zhang2023characterizing} focuses on the bugs and security flaws within Wasm, while others\cite{romano2022wobfuscator,rokicki2022port,oz2023rob,katzman2023gates,puddu2024lack} reveal new attacks based on Wasm. \cite{romano2021empirical} performs both qualitative and quantitative analysis of three open-source Wasm compilers, revealing the characteristics of bugs within these compilers from multiple perspectives. In contrast, \cite{lehmann2020everything} focuses on the execution of Wasm, analyzing its use in browsers, node.js, and server-side standalone runtimes. The study identifies many classic vulnerabilities, which, although mitigated in native binaries, are re-exposed in Wasm. Furthermore, \cite{lehmann2020everything} finds that Wasm introduces some unique attacks. \cite{zhou2023wadiff} and \cite{zhang2023characterizing} focus on detecting bugs in Wasm runtimes. \cite{zhou2023wadiff} designs a differential testing framework called WADIFF, which translates the Wasm specification into a structured domain-specific language (DSL). It then automates test case generation through a symbolic execution engine based on the DSL, and detects bugs by analyzing discrepancies in the test results for the same test case across different runtimes. \cite{zhang2023characterizing} collects bug cases from platforms like GitHub, summarizes the characteristics of these bugs, and develops a pattern-based bug detection framework for automated bug identification. This framework has identified several new bugs across various runtimes, including Wasmtime and Wasmer. \cite{romano2022wobfuscator} presents Wobfuscator, a tool that translates specific JavaScript code into Wasm, thereby evading malware detectors. Experiments show that Wobfuscator can bypass state-of-the-art learning-based static malware detectors without affecting code correctness. For side-channel attacks in browsers, \cite{rokicki2022port} proposes a covert side-channel created through port contention, achieving a bit rate an order of magnitude higher than current state-of-the-art methods. \cite{oz2023rob} introduces a new browser ransomware, RøB, which encrypts victims' local files by exploiting the file system access(FSA) APIs and Wasm, even when access restrictions are applied to the API. Furthermore, \cite{oz2023rob} tests several antivirus software and finds that they fail to detect the malicious behavior of RøB. \cite{katzman2023gates} explores how transient execution can facilitate cache attacks, with a proposed method that works across a wide range of environments, including Wasm. \cite{puddu2024lack} compares two methods for deploying confidential code within the TEE, namely deploying native binaries and deploying IR (Wasm). The experiments show that Wasm-based deployment of confidential code leaks most IR instructions with high accuracy, exposing the confidential code.

\textbf{Secure Runtimes.} Several studies\cite{narayan2021swivel,bosamiya2022provably,johnson2023wave,lei2023put,yavarzadeh2023half,menetrey2023comprehensive} propose various methods to enhance the security of Wasm in response to its security issues. To strengthen Wasm's resistance to Spectre attacks, \cite{narayan2021swivel} introduces the Swivel compiler framework, which improves Wasm security through both software-based methods (Swivel-SFI) and hardware-assisted techniques (Swivel-CET). \cite{bosamiya2022provably} explores two technologies for implementing secure sandboxes, vWasm and rWasm. The former is based on traditional formal methods, while the latter leverages Rust's safety mechanisms. \cite{johnson2023wave} designs a new Wasm runtime, WaVe, which adopts several security strategies to ensure memory safety, file system isolation, and network isolation, outperforming Wasmtime in some benchmarks. \cite{lei2023put} introduces PKUWA, a Wasm runtime with memory isolation. PKUWA overcomes the differences between Wasm's linear memory and MPK pages by using Domain Isolated Linear Memory (DILM), implementing memory protection in Wasm based on MPK, and effectively defending against vulnerabilities such as Heartbleed. In response to branch-based side-channel attacks, \cite{yavarzadeh2023half} proposes a new defense method, Half\&Half, which does not require hardware or ISA modifications but only minor compiler changes. \cite{yavarzadeh2023half} modifies Swivel to implement Half\&Half, reducing the overhead of condition branch isolation in Swivel-SFI by an order of magnitude. \cite{menetrey2023comprehensive} designs TWINE, a runtime for running Wasm applications in TEEs. In addition to supporting extended WASI interfaces, TWINE also provides remote attestation.

In addition, several studies \cite{jiang2023revealing,romano2020wasim,she2024wadec,romano2023function} have focused on the performance, behavioral analysis, and readability of Wasm. \cite{jiang2023revealing} identifies performance issues in server-side Wasm runtimes through differential testing. \cite{romano2020wasim} designs a classifier that uses features of Wasm modules to determine their purpose. Due to the poor readability of Wasm, which makes debugging and analysis challenging, \cite{she2024wadec} leverages fine-tuned LLMs to interpret and decompile Wasm, converting it into more understandable C code. \cite{romano2023function} reveals issues with inline assembly optimizations in Wasm compilers, which slow down the execution of Wasm modules.

\textbf{Breaking Resource Isolation.} Several studies have investigated threats to resource availability in containerized and virtual machine (VM) environments \cite{gao2019houdini,xiao2023attacks,yang2021demons,sharma2016containers}. These works reveal that flaws in the implementation of isolation mechanisms can be exploited to consume host resources, breaking resource isolation mechanisms. \cite{sharma2016containers} compares container and VM performance in large-scale data centers, noting that containers, while lightweight and efficient, provide weaker isolation in multi-tenant settings. The cgroup mechanism controls container resource usage (CPU, memory, I/O), but \cite{gao2019houdini} shows that attackers in Docker containers can bypass it by injecting workloads into OS components (e.g., upcalls, services, interrupts), leading to excessive host resource consumption. Furthermore, \cite{yang2021demons} introduces abstract resource attacks on shared kernel-level data structures, identifying many exploitable targets via static analysis and showing their significant impact. VM technology offers stronger isolation than containers by using separate OS kernels. To combine VM-level isolation with container efficiency, Kata Containers and Firecracker adopt lightweight VM-based designs. Nevertheless, \cite{xiao2023attacks} shows that attackers can still exhaust host resources via operation forwarding, where guest OS operations propagate through container runtimes, device emulators, or host kernel components.

Previous works \cite{gao2019houdini,xiao2023attacks,yang2021demons,sharma2016containers} have primarily focused on resource exhaustion issues in container and virtual machine environments, whereas our work targets such issues in the context of Wasm runtimes. First, we measure the attack surface in WASM runtimes. Then, we use the attack ideas from prior studies to design the corresponding attack strategies for WASM runtimes to verify their effectiveness on WASM runtimes. More importantly, we identify new attack surfaces of Wasm runtimes that have not been analyzed in prior work. Based on these findings, we design new attack strategies and demonstrate their effectiveness through experiments. In addition, we design attack strategies that combine multiple interfaces to perform more complex and serious attacks. Furthermore, to perform static analysis in WASM runtimes, we propose a new static analysis method specifically designed for Rust. For indirect call sites related to Rust features, our method can provide more precise analysis results, laying a solid foundation for subsequent analysis. Moreover, our analysis approach can also be applied to other Rust projects.

\section{Conclusion}

Wasm runtimes serve as secure containers, enabling the execution of Wasm modules outside of browsers, with WASI/WASIX interfaces to provide system services. Although these runtimes claim to run Wasm modules securely and efficiently, we observe that it is possible to exploit these WASI/WASIX interfaces, breaking the resource isolation and exhausting host resources. In this paper, we propose several static approaches to systematically explore the resource attack surface of Wasm runtimes. Then, several exploitation strategies are proposed, targeting file systems, I/O, and network interfaces that can be used by Wasm instances to bypass the isolation of standalone runtimes. Through detailed experiments, we demonstrated the effectiveness of these strategies and assessed the impact on the host system. Additionally, we discuss the mitigation strategies to limit the impact of the attacks.

\section*{Acknowledgments}

This work was supported by the National Natural Science Foundation of China under Grants No. 62302122 and No. 62172123, the Key R\&D Program of of Heilongjiang Province of China under Grants No. JD2023SJ07, and CCF-Huawei Populus Grove Fund under Grants No. CCF-HuaweiSY202411.

\section{Open science}

In alignment with the USENIX Security open science policy, we are committed to enhancing the reproducibility and replicability of our research findings. To support this, we will make the artifacts related to our work publicly available on \href{https://github.com/zhaofengyu-hit/wasm_analyzer_exploit.git}{Github}, including the source code, scripts, binaries, and other materials necessary for replicating our experiments. By sharing these artifacts, we aim to facilitate the reproduction of our results and encourage further exploration and validation by the research community. Our artifact is also available through Zenodo at \href{https://doi.org/10.5281/zenodo.15597665}{10.5281/zenodo.15597665}. In accordance with the policy, we will ensure that the Artifact Evaluation committee has access to the relevant materials after paper acceptance and before the final papers are due.

\section{Ethical considerations}

WebAssembly (Wasm) has been increasingly adopted in a wide range of scenarios. However, we have identified that some Wasm runtimes exhibit insufficient resource isolation capabilities. Our goal is to systematically explore the attack surface related to resource isolation in Wasm runtimes, with the aim of raising awareness of potential security risks and promoting the development of related defense techniques. Ultimately, we hope our work contributes to enhancing the security of cloud environments. To this end, we propose a static analysis framework for analyzing Wasm runtimes and uncovering potential security issues. Throughout the research and experimentation process, we strictly adhered to ethical principles to ensure responsible conduct.

First, our study focuses exclusively on publicly available, non-sensitive software. We did not use any proprietary or private data. Second, to avoid any negative impact on real-world systems, all experiments were conducted in a fully controlled local testbed that was not connected to any external devices or networks, thereby ensuring that no external systems or users could be affected. Finally, all scripts, source code, and binary files used for analysis are shared in a reproducible manner, allowing the research community to replicate and build upon our work.

Our findings are relevant to five categories of stakeholders: (1) Wasm runtime developers and maintainers; (2) developers of applications that rely on Wasm runtimes; (3) end users of Wasm-based applications; (4) cloud service providers and tenants; and (5) the security research and defense community. For runtime developers, the proposed attack surface may affect their products and user experience, and addressing these issues may incur engineering costs and reputational concerns. Application developers may face risks if their systems rely on vulnerable runtime interfaces that cannot be mitigated through conventional configurations. For end users, malicious applications could degrade device performance, resulting in denial-of-service scenarios and a poor user experience. Cloud providers that use Wasm runtimes as execution environments may be vulnerable to resource competition attacks by malicious tenants, reducing service quality for normal users. For the security research and defense community, our findings can inspire new research on exploitation and defenses. Improper disclosure might risk malicious use, so responsible disclosure is important.

Our intention is to identify and disclose potential attack surfaces to support remediation and to foster further research in isolation and defense mechanisms. To minimize impact on stakeholders, we have taken several precautionary steps. We conducted responsible disclosure by contacting relevant runtime security teams, providing detailed technical descriptions and suggested remediation steps. We have discussed security risks and mitigation approaches with these security teams, and we will continue to work with them to explore more appropriate protection measures. For stakeholders we could not contact directly (e.g., application developers, users, and cloud service providers), we have advised runtime developers to document the risks and provide some mitigation approaches. At the same time, we recognize that our work can be misused when open-sourced. For instance, our static analysis approach could be used to discover vulnerabilities for malicious purposes. To address this, we will explicitly prohibit malicious use when releasing any related tools or code.

As discussed in Section 6, we propose several mitigation strategies along with the corresponding analysis. First, we can build defenses based on existing tools. The advantage of this defense strategy is the low implementation cost, but it is difficult to fully defend against the proposed attacks. Besides, monitoring all types of host resources and accurately identifying the resource consumers could enhance security but would introduce high performance overhead. A possible trade-off is to focus monitoring efforts on a subset of sensitive resources, aiming to balance both security and performance.

Our findings contribute to a better understanding of the attack surface in Wasm runtimes and can help promote the development of more secure isolation mechanisms. The paper includes detailed analysis and mitigation suggestions. We are aware of the potential misuse risks and therefore adopted a responsible disclosure process to ensure that the information is used constructively. We hope our research can motivate future work in this area, encouraging others to explore new defenses while upholding the same ethical standards.

\printbibliography

\end{document}